\documentclass[journal,twoside, a4paper]{IEEEtran}
\usepackage{multicol}
\usepackage{makecell}
\usepackage{color}
\usepackage{amssymb}
\usepackage{rotating}
\usepackage[T1]{fontenc}
\usepackage{lscape}
\usepackage{dblfloatfix}
\usepackage{xcolor}
\usepackage{subcaption}
\usepackage[ruled,vlined]{algorithm2e}
\usepackage[autostyle]{csquotes}
\usepackage{tikz}
\usepackage{acro}
\usepackage{siunitx}
\usepackage{enumitem}
\usepackage{multirow}
\usepackage{cite}
\usepackage{hyperref}
\usepackage{amsmath,amssymb,amsfonts}
\usepackage{tikz}
\usepackage{float}
\usepackage{algorithmic}
\usepackage{graphicx}
\usepackage{textcomp}
\usepackage{xcolor}
\usepackage{longtable}
\usepackage{url}
\usepackage{booktabs}
\usepackage{comment}
\def\BibTeX{{\rm B\kern-.05em{\sc i\kern-.025em b}\kern-.08em
    T\kern-.1667em\lower.7ex\hbox{E}\kern-.125emX}}

\ifCLASSINFOpdf
\else
 
\fi

\usepackage[a4paper, total={180mm,257mm}]{geometry}

\begin{document}

\title{Harnessing Transformers: A Leap Forward in Lung Cancer Image Detection}

\author{\IEEEauthorblockN{
Amine Bechar\IEEEauthorrefmark{1}, 
Youssef Elmir\IEEEauthorrefmark{1}, 
Rafik Medjoudj\IEEEauthorrefmark{1},
Yassine Himeur\IEEEauthorrefmark{2}and
Abbes Amira\IEEEauthorrefmark{3,4} 
}\\
\IEEEauthorblockA{\IEEEauthorrefmark{1}
Laboratoire LITAN École supérieure en Sciences et Technologies de l’Informatique et du Numérique RN 75, Amizour
06300, Bejaia, Algérie (bechar@estin.dz; elmir@estin.dz; medjoudj@estin.dz)}\\

\IEEEauthorblockA{\IEEEauthorrefmark{2}College of Engineering and Information Technology, University of Dubai, Dubai, UAE (yhimeur@ud.ac.ae)}\\
\IEEEauthorblockA{\IEEEauthorrefmark{3}Department of Computer Science, University of Sharjah Sharjah, UAE
(aamira@sharjah.ac.ae)}\\
\IEEEauthorblockA{\IEEEauthorrefmark{3}Institute of Artificial Intelligence, De Montfort University Leicester, United Kingdom\\
}}

\maketitle

\begin{abstract}
This paper discusses the role of Transfer Learning (TL) and transformers in cancer detection based on image analysis. With the enormous evolution of cancer patients, the identification of cancer cells in a patient’s body has emerged as a trend in the field of Artificial Intelligence (AI). This process involves analyzing medical images, such as Computed Tomography (CT) scans and Magnetic Resonance Imaging (MRIs), to identify abnormal growths that may help in cancer detection. Many techniques and methods have been realized to improve the quality and performance of cancer classification and detection, such as TL, which allows the transfer of knowledge from one task to another with the same task or domain. TL englobes many methods, particularly those used in image analysis, such as transformers and Convolutional Neural Network (CNN) models trained on the ImageNet dataset. This paper analyzes and criticizes each method of TL based on image analysis and compares the results of each method, showing that transformers have achieved the best results with an accuracy of 97.41\% for colon cancer detection and 94.71\% for Histopathological Lung cancer. Future directions for cancer detection based on image analysis are also discussed.

\end{abstract}


\IEEEpeerreviewmaketitle

\section{Introduction}
The application of artificial intelligence (AI) and image analysis is revolutionizing various sectors of healthcare \cite{sayed2023time,himeur2023ai}. Deep learning enables computers to interpret medical images such as X-rays and MRI scans, offering automatic detection of issues like tumors and fractures \cite{copiaco2023innovative,habchi2023ai}. AI can detect patterns and discrepancies in images, aiding in precise diagnoses at a faster rate than humans. This speed allows doctors to focus on critical cases \cite{himeur2023face}. AI also merges image data with genetic details to offer personalized medical treatment \cite{ref3,ref4}. In 2022, the U.S. witnessed approximately 1.9 million new cancer cases and over 600,000 related deaths \cite{ref5}. Efforts to curb this ailment have given rise to methods combining image analysis and AI, especially given the influx of medical data \cite{ref6,boughorbel2022predicting}. AI, particularly in breast cancer screenings, has proven valuable in enhancing cancer detection, aiding in distinguishing between benign and malignant tumors \cite{ref8,hamza2023hybrid}.

Handling vast medical imaging data can be a challenge due to time constraints, privacy, and security concerns, especially with limited labeled data, impacting the training of machine learning models \cite{sayed2023time}. However, advancements in machine learning, like transfer learning (TL), help navigate these challenges \cite{sayed2022deep,sohail4348272deep}. TL reuses knowledge from one task to enhance the performance on another related task \cite{ref9,copiaco2022exploring}. This method has been pivotal in cancer detection, including colon \cite{ref10}, skin \cite{ref11}, and breast cancers \cite{ref12}. TL capitalizes on extensive datasets like ImageNet for tasks where labeled data is scarce, such as medical imaging, addressing the diverse features needed to discern between different cancers. This work highlights that, despite limited data, TL and transformers excel in cancer classification through medical imagery.

This paper advances the field of medical image analysis in several ways. It conducts a comparative study of transfer learning models for classifying images of various cancers like lung, colon, and breast. This research provides insights into which models excel in specific imaging domains for cancer detection. The paper introduces a method using ViTs for cancer detection, detailing each stage from data acquisition to classification. Notably, using ViTs yields promising cancer detection results, underscoring their potential in this field. The paper also gives guidance on choosing models for medical image analysis based on cancer dataset traits and model features.

\section{Related works}
In recent years, several studies have proposed models and metrics for cancer detection using machine learning and computer vision algorithms. For example, Duong et al \cite{ref13} developed a solution for tuberculosis detection from chest X-ray images using EfficientNet and ViT, achieving an accuracy of 97\%. However, they did not provide specific information about dataset sizes and potential biases, which could affect the generalizability of the results. Zidan et al \cite{ref14} addressed the lack of labeled data in colorectal cancer by proposing SwinCup, a transformer-based model with a hierarchical Swin Transformer encoder-decoder architecture for accurate segmentation, achieving an F1 score, recall, and precision of 0.92. Nevertheless, further evaluation of a wider range of datasets is needed to establish the model's generalizability.

\begin{table*}[t!]
\caption{Comparative study of the related works.}
\label{tab:comparative_study}
\scriptsize
\begin{tabular}{
m{5mm} 
m{5mm}
m{40mm}
m{30mm}
m{30mm}
m{12mm}
m{20mm}
}
\hline
\textbf{Ref.} & \textbf{Year} & \textbf{Contribution} & \textbf{TL model} & \textbf{Name of dataset} & \textbf{N° of images} & \textbf{Best metric} \\ \hline
\cite{ref13} & 2021 & A practical framework for the accurate detection of tuberculosis from chest X-ray images. & EfficientNet and ViT & Montgomery County CXR & 3955 & Accuracy=97.72\% \\ \hline
\cite{ref14} & 2023 & Addressing the segmentation of histopathological structures in colorectal cancer. & SwinCup & GlaS & 165 & F1=92\% \\ \hline
\cite{ref15} & 2023 & Training complex architectures from scratch on histopathology tasks & ConvNeXt & PCam & 262,144 & Accuracy=90.31\% \\ \hline
\cite{ref16} & 2022 & Proposing a novel method, for breast ultrasound detection using ViTs & BUViTNet & Mendeley & 780 & AUC=96.8\% \\ \hline
\cite{ref17} & 2023 & Applying TL and transformers in the field of medical imaging & Inception-V3 & CheXpert & 224,316 & F1=64\% \\ \hline
\cite{ref18} & 2022 & Introducing the use of state-of-the-art ViTs for kidney disease diagnosis & Swin Transformer & CT KIDNEY & 12,446 & Accuracy=99.30\% \\ \hline
\end{tabular}
\end{table*}

Springenberg et al \cite{ref15} proposed a solution to address the issue of low performance and instability when training complex architectures from scratch on histopathology tasks. They suggest utilizing TL with CNN models ConvNeXt-L, GasHis, and Inception-v3, which have been pretrained on the ImageNet large-dataset. The ConvNeXt-L model achieved the best results with an accuracy of 0.90 and AUC of 0.972. However, the authors did not explore the limitations and challenges that these models may face when applied in real-world scenarios. Ayana et al \cite{ref16} proposed a solution for accurate classification of breast ultrasound images for the detection of breast cancer using ViTs  pretrained on the ImageNet dataset. The authors compared the performance of their proposed method, BUViTNet, with other models and evaluated it using metrics such as accuracy. The proposed method achieved high scores, including an AUC of 1 ± 0 and MCC score of 1 ± 0 on the Mendeley dataset. However, the study's main limitation was the small sample size used for training the ViT models.

Usman et al. \cite{ref17} utilized ViT for medical imaging on the CheXpert dataset and compared its performance to CNN architectures like VGG-16 and ResNet. ViT had the highest F1 score of 0.57, but a comprehensive metric analysis was not presented. Meanwhile, Islam et al. \cite{ref18} aimed to enhance kidney disease diagnosis accuracy from CT scan images using transfer learning and the Swin Transformer. This model achieved an impressive 99.30\% accuracy. However, its performance was only tested on their dataset, lacking external validation. A comparison of these studies is presented in Table 1.

\section{Proposed methodology}
The proposed methodology for using ViT in cancer classification and detection is a multistage process that involves getting, preparing, and analyzing data. Before starting the preparation stage, it is important to identify the type of images, format, and sources. This ensured that the collected data were suitable for training and evaluating the model. Figure 1 shows the steps of the proposed methodology.

\subsection{Used Datasets}
The first step in the proposed methodology is to use a number of datasets that are sufficiently large to provide sufficient data for training and evaluating the model. The images should be of high quality and should be labelled with the appropriate classification (malignant or benign). The datasets used are presented in section 4.1.

\subsection{Data preprocessing}
The cancer dataset needs to be pre-processed to ensure that it is suitable for training the model. This may include tasks such as resizing, normalization, and data cleaning. The goal of data preprocessing is to ensure that the dataset is consistent and free of errors.

\subsection{Data augmentation}
Data augmentation (DA) techniques can be used to increase the size of the dataset, improve the performance of the model, and solve the problem of imbalanced classes. This step may include tasks, such as rotation, flipping, and cropping. The goal of DA is to provide a model with more examples of the same type of data, which can help it learn more effectively.
\begin{figure*}[t!]
    \begin{center}
    \includegraphics[width=1\textwidth]{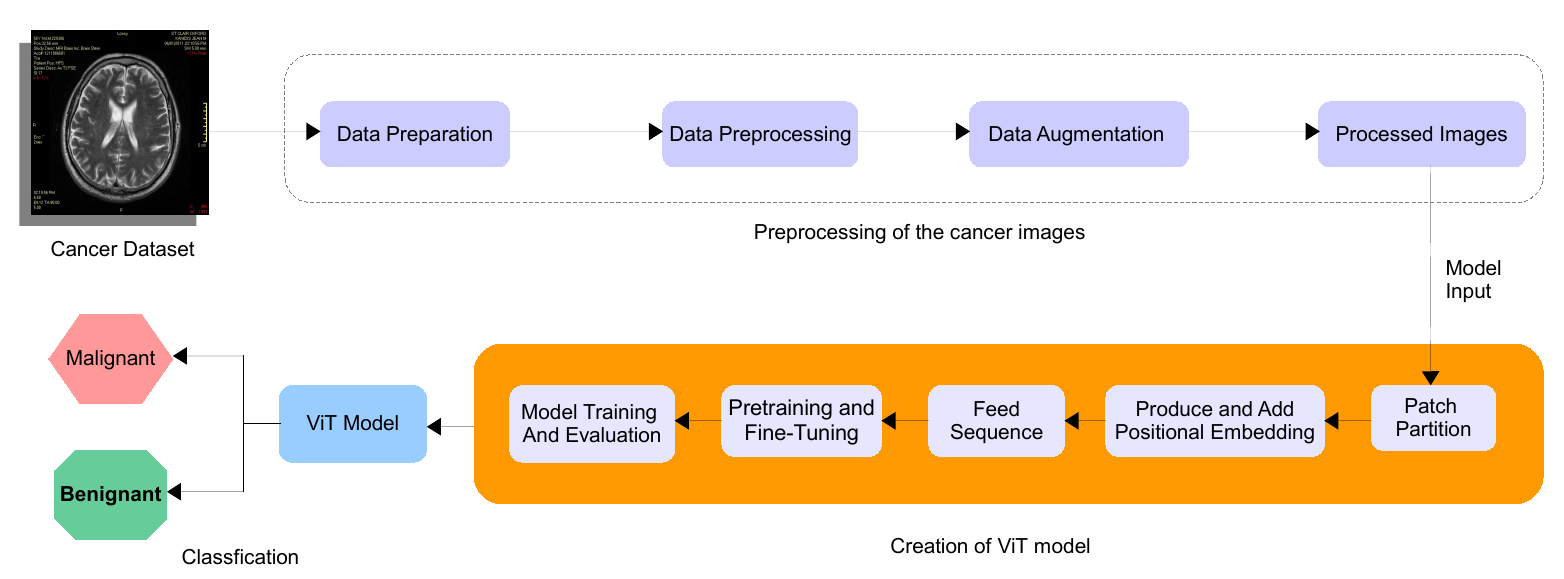}\\
    \end{center}
    \caption{ViT architecture for cancer detection}
    \label{fig:HFL}
\end{figure*}

\subsection{Creation of the ViT model}
The proposed methodology uses the ViT model for cancer classification and detection. This involves the following steps:

\begin{itemize}
\item Patch partition: The image was divided into smaller patches of equal size. This step is necessary because the ViT model treats images as sequences of patches \cite{ref19}.
\item Produce and add positional embedding: Positional embeddings are added to patch embeddings to provide information about the spatial location of each patch. This step is important because it allows the model to learn spatial relationships between patches \cite{ref19}.
\item Feed sequence: Patch embeddings with positional embeddings were fed into a standard transformer encoder. This step is where the actual learning takes place, as the transformer encoder processes the sequence of patch embeddings and positional embeddings.

\item Multi-head attention:  Each patch attends to all other patches in the image. This allows the model to capture global dependencies and relationships between different regions of the image. The outputs of the attention heads are then combined and processed further to generate the final representation of the image \cite{ref20}.
\item Multi-layer perceptron: MLP works by passing input data through multiple layers of interconnected nodes (neurons) to perform a classification task. Each neuron in a layer receives inputs from the previous layer, applies a weighted sum of the inputs, and passes the result through an activation function to produce an output. This output is then propagated to the next layer as input \cite{ref20}.
 \item Pretraining and fine-tuning: The model was pretrained on a large dataset (ImageNet) and then fine-tuned on a smaller dataset of images specific to the target task, such as cancer classification. This allows the model to use the knowledge and learn features relevant to the target task, which make the training faster and more accurate. 
\item Training and evaluation: The performance of the model is evaluated on a separate test dataset to assess its accuracy and other performance metrics. This step is important because it allows us to determine how well the model is performing on the target tasks.     

\end{itemize}

\subsection{Classification}
The trained model can be used to classify new images of malignant or benignant cancer. This step is the ultimate goal of the proposed methodology, as it allows us to classify new images based on the features learned by the model. Algorithm 1 explains in detail ViT in the proposed methodology.

\begin{figure*}[t!]
\centering
\includegraphics[width=1\textwidth]{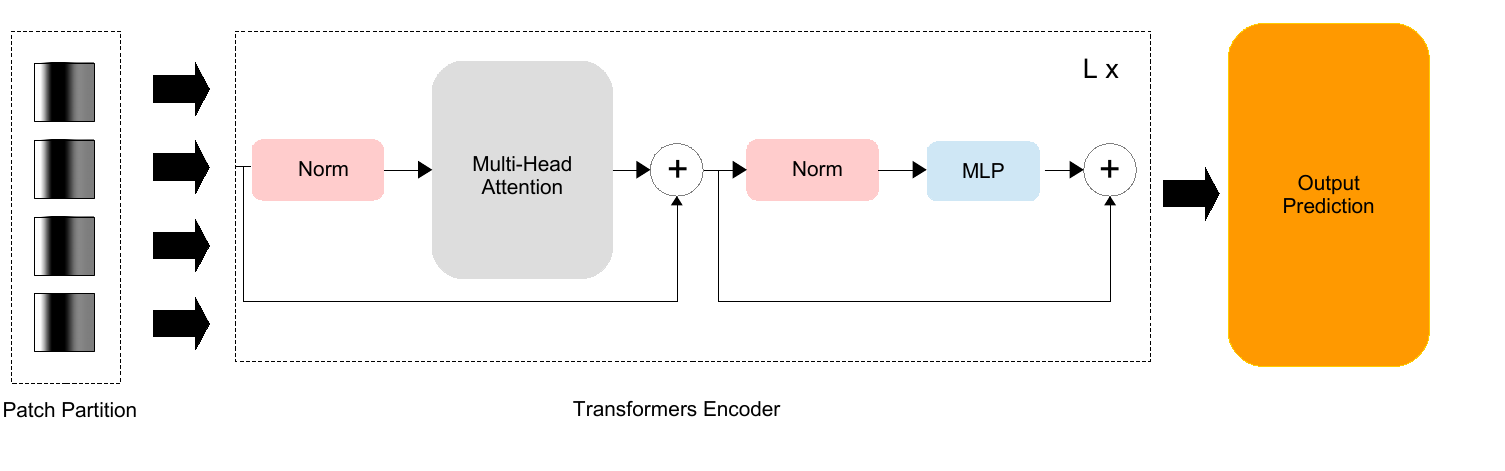}\\
\caption{Detailed steps of Transformer Encoder \cite{ref20}.}
\label{fig:HFL}
\end{figure*}

\section{Results}
This section presents the results obtained by applying TL techniques to cancer datasets and compares them with those of the proposed methodology to determine their effectiveness. In addition, the datasets used in the experiments are described, and the output results are presented and discussed.

\subsection{Datasets}
In this study, five datasets of five types of malignant and benign cancer (brain, breast, lung, colon, and skin cancer) were used. These datasets comprise images in the JPEG format, including pictures, ultrasound images, and histopathological images. All of these datasets have been created and generated for classification and cancer detection in academic research. The following sections provide a detailed description of each dataset, including its characteristics, data collection methods, and features. Table 2 describes the datasets used in the experiments.

\begin{itemize}
\item Lung and Colon Cancer Histopathological Images: The dataset consisted of 25,000 histopathological images of lung and colon cancer. Within the dataset, there were three distinct classes of lung cancer and two classes of colon cancer. Each image was measured at $768 \times 768$ pixels and saved in JPEG file format. The dataset was generated through the acquisition of 750 distinct images of lung tissue, comprising 250 benign lung tissue samples, 250 adenocarcinomas, and 250 carcinomas. All images we obtained from pathology glass slides \cite{ref21}.

\item Brain Tumor Dataset: The brain tumor dataset contained 3064 T1-weighted contrast-enhanced images from 233 patients with three types of brain tumors: Meningioma (708 slices), Glioma (1426 slices), and Pituitary tumor (930 slices). The images had an in-plane resolution of $512 \times 512$ pixels with a pixel size of 0.49 × 0.49 mm2. The slice thickness was 6 mm, and the slice gap was 1 mm. This data was used in \cite{ref22} and \cite{ref23}.

\item Skin cancer HAM10000: The HAM10000 dataset is a collection of 10,015 dermatoscopic images of skin lesions that can be used to train the ML models. The dataset includes images from different populations acquired using different modalities, providing a diverse set of images for training. The images covered all important diagnostic categories of pigmented lesions, including melanoma, basal cell carcinoma, and vascular lesions. The dataset is publicly available and can be used for academic machine-learning purposes \cite{ref24}. 

\item CBIS-DDSM: Breast Cancer Image Dataset is a collection of mammograms from various sources, including Massachusetts General Hospital, Wake Forest University School of Medicine, and Sacred Heart Hospital \cite{ref25}, which contains 2,620 scanned film mammography studies, including normal, benign, and malignant cases, with verified pathology information. Figure 3 represents a sample of the dataset.
\end{itemize}

\begin{algorithm*}
\caption{Description of the proposed ViT Algorithm.}
\label{alg:ViT}

\textbf{Input:} Image Class\\
\textbf{Output:} Class label\\
\textbf{Initialization:}\\
//\textit{Input Processing}

$patches \gets \text{SplitImage(Inputimage)}$\;

$flattenedPatches \gets \text{FlattenPatches(patches)}$\;

$lowerDimEmbeddings \gets \text{LinearProjection(flattenedPatches)}$\;

$patchesWithPosition \gets \text{AddPositionalEmbeddings(lowerDimEmbeddings)}$\;

//\textit{Transformer Encoder}

$sequence \gets \text{FeedSequence(patchesWithPosition)}$\;

$attention \gets \text{MultiHeadAttention(sequence)}$\;

$feedForward \gets \text{FeedForwardLayer(attention)}$\;

//\textit{Transfer Learning}

$preTrainedModel \gets \text{PretrainModel(ImageNet)}$\;

$fineTunedModel \gets \text{FineTuneModel(preTrainedModel, task)}$\;

//\textit{Output Prediction}

$output \gets \text{OutputPrediction(feedForward)}$\;
\Return{$output$}\;

\end{algorithm*}

\subsection{Illustration of the results}
This section presents the important findings of this study, which are organized into three subsections. The experimental setup and evaluation metrics were described, and the obtained results were visually represented using tables and figures. This section provides a comprehensive overview of the study and highlights the key findings.

\begin{figure*}[t]
\begin{subfigure}{.5\textwidth}
  \centering
  \includegraphics[width=.8\linewidth,height=160pt]{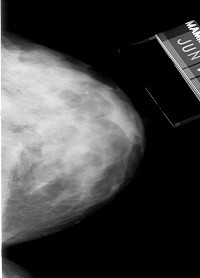}
  \caption{Malignant breast tumor}
  \label{fig:sub1}
\end{subfigure}%
\begin{subfigure}{.5\textwidth}
  \centering
  \includegraphics[width=.8\linewidth,height=160pt]{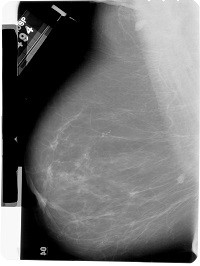}
  \caption{Benign breast tumor}
  \label{fig:sub2}
\end{subfigure}
\caption{Example of sample of CBIS-DDSM Cancer dataset}
\label{fig:test}
\end{figure*}

\begin{figure*}[t]
\begin{subfigure}{.5\textwidth}
  \centering
  \includegraphics[width=.8\linewidth]{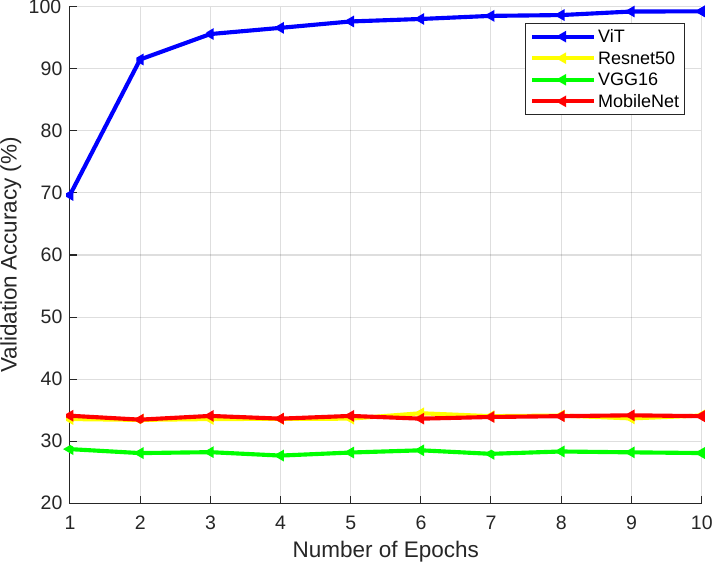}
  \caption{Validation Accuracy}
  \label{fig:sub1}
\end{subfigure}%
\begin{subfigure}{.5\textwidth}
  \centering
  \includegraphics[width=.8\linewidth]{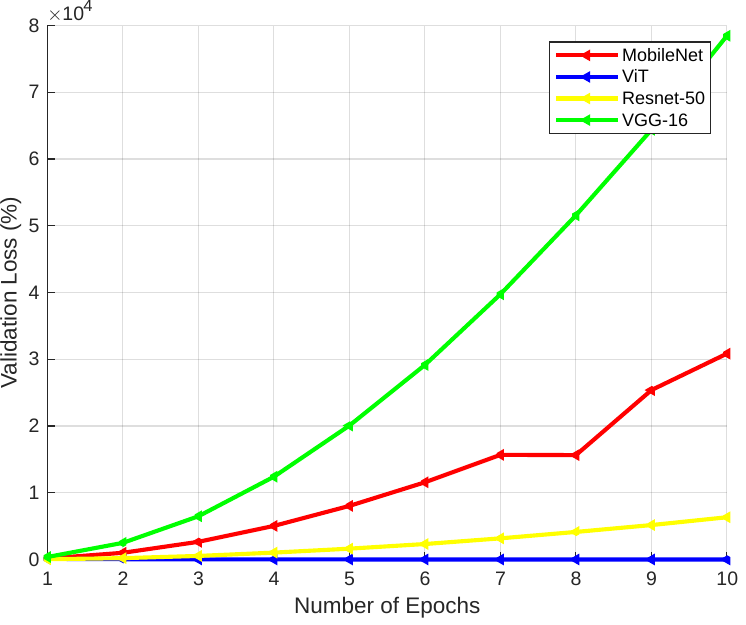}
  \caption{Validation Loss}
  \label{fig:sub2}\
\end{subfigure}
\caption{Results of applying TL models on Lung Cancer dataset}
\label{fig:test}
\end{figure*}

\begin{table*}[t!]
\caption{Description of features of the used datasets.}
\captionsetup[table]{labelfont=bf,textfont=it,labelsep=period,skip=10pt}
\label{tab:example}
\renewcommand{\arraystretch}{1.5}
\begin{tabular}{@{}p{6cm}cccccc@{}}
\toprule
\hline
 & \textbf{Lung H. Images} & \textbf{Colon H. Images} & \textbf{Skin HAM10000} & \textbf{CBIS-DDSM} & \textbf{Brain Tumor} \\ \midrule
 \hline
\#Class & 3 & 2 & 7 & 3 & 2 \\

\#Number of Samples & 15000 & 10000 & 10000 & 10239 & 10000  \\

\#Train & 12000 & 8000 & 8000 & 8191 & 8000 \\

\#Test & 3000 & 2000 & 2000 & 2048 & 2000  \\
\hline
\end{tabular}
\end{table*}

\subsubsection{Experimental setup}
Experiments were conducted on a machine equipped with an Intel Core i7 CPU and NVIDIA GeForce RTX 3080 GPU. Python 3.10.1 and TensorFlow 2.8.0 were used, except for ViT, which employed Pytorch 2.2.0. Standard data augmentation techniques, such as random cropping, horizontal flipping, and normalization, were applied to the training set. The datasets discussed earlier were split into training, validation, and test sets at a ratio of 80:10:10.

The CNN models were pre-trained on the ImageNet dataset to leverage the knowledge acquired from this large-scale dataset. Each model was subsequently fine-tuned on the training set for 10 epochs to avoid the risk of overfitting. A batch size of 75 was employed for all the experiments. The Adam optimizer with a learning rate of 0.001 and categorical cross-entropy loss function were utilized.

\begin{table*}[t!]
\caption{Validation accuracy and loss of TL models on cancer datasets.}
\label{tab03}
\small
\begin{tabular}{
    m{20mm}
    @{\hspace{6pt}}m{15mm}@{\hspace{6pt}}
    @{\hspace{6pt}}m{10mm}@{\hspace{6pt}}
    @{\hspace{6pt}}m{10mm}@{\hspace{6pt}}
    @{\hspace{6pt}}m{10mm}@{\hspace{6pt}}
    @{\hspace{6pt}}m{10mm}@{\hspace{6pt}}
    @{\hspace{6pt}}m{10mm}@{\hspace{6pt}}
    @{\hspace{6pt}}m{10mm}@{\hspace{6pt}}
    @{\hspace{6pt}}m{10mm}@{\hspace{6pt}}
    @{\hspace{6pt}}m{10mm}@{\hspace{6pt}}
    @{\hspace{6pt}}m{20mm}@{\hspace{6pt}}
}
\hline
Datasets & \multicolumn{2}{c}{Brain Tumor} & \multicolumn{2}{c}{Colon H. Images} & \multicolumn{2}{c}{Skin HAM10000} & \multicolumn{2}{c}{CBIS-DDSM} & \multicolumn{2}{c}{Lung H. Images} \\
Models & Acc (\%) & Loss (\%) & Acc (\%) & Loss (\%) & Acc (\%) & Loss (\%) & Acc (\%) & Loss (\%) & Acc (\%) & Loss (\%) \\
\hline
Transformers & 95.53 & 159.66\vspace{2pt} & 97.41 & 153.2\vspace{2pt} & 83.22 & 43.52\vspace{2pt} & 97.09 & 122.75\vspace{2pt} & 94.47 & 200.38 \\
VGG-16 & 96.32 & 139.15\vspace{2pt} & 99.80 & 6.02\vspace{2pt} & 83.22 & 43.52\vspace{2pt} & 85.56 & 41.51\vspace{2pt} & 28.2 & 136107.73 \\
Resnet-50 & 96.62 & 119.83\vspace{2pt} & 76.71 & 498.88\vspace{2pt} & 85.71 & 41.07\vspace{2pt} & 97.94 & 55,126\vspace{2pt} & 31.32 & 245219.25 \\
MobileNet & 96.06 & 322.63\vspace{2pt} & 99.85 & 15.53\vspace{2pt} & 84.71 & 42.06\vspace{2pt} & 73.51 & 857.68\vspace{2pt} & 33.91 & 30509.04 \\
\hline
\end{tabular}
\end{table*}
\subsubsection{Evaluation metrics} 
The performance of the CNN models was evaluated using validation accuracy, which measures the proportion of correctly classified samples. These metrics were employed due to their general use in image classification tasks and their ability to provide a reliable evaluation of model performance. \textcolor{black}{The formula of validation accuracy is:}

\textcolor{black}{Accuracy = $ \frac{(TP + TN)}{(TP + TN + FP + FN)}  $ Where:}

\begin{multicols}{2}
\begin{itemize}
    \item \textcolor{black}{TP = True Positives}
    \item \textcolor{black}{TN = True Negatives}
    \item \textcolor{black}{FP = False Positives}
    \item \textcolor{black}{FN = False Negatives}
\end{itemize}
\end{multicols}

\subsubsection{Presentation of results}
The results of the comparative study of TL models on four datasets are presented. 
Table 3 displays the accuracy of each CNN model on the CBIS-DDSM dataset. It can be observed that the highest and almost identical accuracy and lowest loss were achieved by ResNet and ViT, while VGG16 and MobileNet had the lowest accuracy and highest loss. VGG16 and MobileNet achieved almost 100\% accuracy on the Colon Cancer Histopathological Images dataset, indicating a potential overfitting problem. ViT demonstrated the best accuracy and lowest loss, suggesting their robustness for this dataset. ResNet performed the lowest accuracy and highest loss, indicating it may not be the optimal choice for this particular dataset.

For the Skin HAM10000 dataset, all CNN models and ViT achieved similar performance in terms of accuracy and loss validation. ResNet achieved the best results, while the proposed ViT performed the worst results compared to the CNN models. On the Brain Cancer dataset, ResNet achieved the highest accuracy, followed by VGG16 and MobileNet, which demonstrated good performance with similar accuracy and loss. ViT performed the lowest accuracy and the highest loss. 
On the Lung dataset with 3 classes, Figure 4 shows that ViT achieved the best results, while VGG16, ResNet, and MobileNet performed poorly, suggesting that these models may not be suited for this particular dataset.

\subsection{Discussion}
This study compared the performance of Transformers and CNNs on five datasets, revealing Transformer's superior ability to manage large datasets \textcolor{black}{effectively due to the MSA mechanism in ViTs which enables more robust learning of global contexts and spatial relationships.} This appeared to be crucial for handling the complexity and high dimensionality of histopathology and radiology images. While CNNs struggled with subtle image features, especially in medical images, Transformers showcased more robust performance. The findings emphasize the importance of model choice based on dataset specifics. Although CNNs are popular for image classification, Transformers might be better for medical image analysis. However, the study only evaluated limited models and datasets and solely used accuracy as the metric. Future research could diversify models, datasets, and metrics like sensitivity and F1 score.

\section{Conclusion}
This research meticulously compared two prominent machine learning models: Transformers and Convolutional Neural Networks (CNNs) across five distinct datasets. Our comprehensive analysis illuminated the consistent superiority of Transformers over CNNs in all datasets tested. Notably, Transformers displayed remarkable precision, making them a promising avenue for image classification tasks, particularly in the nuanced field of medical images for cancer detection. This is not to wholly discredit the capabilities of CNNs, which have their merits and have been pivotal in many image classification tasks over the years. However, the findings of this study underscore the potential of Transformers as a robust tool that may very well shape the future of image-based cancer detection techniques. The insights gleaned from this research serve as a beacon for researchers and medical professionals alike. By providing clear comparative data, this study enables an informed choice of models tailored to specific tasks and datasets. %As we look to the future, these findings not only bolster the utility of Transformers in medical image analysis but also pave the way for further exploration, optimization, and potential breakthroughs in the realm of AI-driven medical diagnostics.



\begin{thebibliography}{10}
\providecommand{\url}[1]{#1}
\csname url@samestyle\endcsname
\providecommand{\newblock}{\relax}
\providecommand{\bibinfo}[2]{#2}
\providecommand{\BIBentrySTDinterwordspacing}{\spaceskip=0pt\relax}
\providecommand{\BIBentryALTinterwordstretchfactor}{4}
\providecommand{\BIBentryALTinterwordspacing}{\spaceskip=\fontdimen2\font plus
\BIBentryALTinterwordstretchfactor\fontdimen3\font minus \fontdimen4\font\relax}
\providecommand{\BIBforeignlanguage}[2]{{%
\expandafter\ifx\csname l@#1\endcsname\relax
\typeout{** WARNING: IEEEtran.bst: No hyphenation pattern has been}%
\typeout{** loaded for the language `#1'. Using the pattern for}%
\typeout{** the default language instead.}%
\else
\language=\csname l@#1\endcsname
\fi
#2}}
\providecommand{\BIBdecl}{\relax}
\BIBdecl

\bibitem{sayed2023time}
A.~N. Sayed, Y.~Himeur, and F.~Bensaali, ``From time-series to 2d images for building occupancy prediction using deep transfer learning,'' \emph{Engineering Applications of Artificial Intelligence}, vol. 119, p. 105786, 2023.

\bibitem{himeur2023ai}
Y.~Himeur, M.~Elnour, F.~Fadli, N.~Meskin, I.~Petri, Y.~Rezgui, F.~Bensaali, and A.~Amira, ``Ai-big data analytics for building automation and management systems: a survey, actual challenges and future perspectives,'' \emph{Artificial Intelligence Review}, vol.~56, no.~6, pp. 4929--5021, 2023.

\bibitem{copiaco2023innovative}
A.~Copiaco, Y.~Himeur, A.~Amira, W.~Mansoor, F.~Fadli, S.~Atalla, and S.~S. Sohail, ``An innovative deep anomaly detection of building energy consumption using energy time-series images,'' \emph{Engineering Applications of Artificial Intelligence}, vol. 119, p. 105775, 2023.

\bibitem{habchi2023ai}
Y.~Habchi, Y.~Himeur, H.~Kheddar \emph{et~al.}, ``Ai in thyroid cancer diagnosis: Techniques, trends, and future directions,'' \emph{arXiv preprint}, vol. arXiv:2308.13592, 2023.

\bibitem{himeur2023face}
Y.~Himeur, S.~Al-Maadeed, I.~Varlamis \emph{et~al.}, ``Face mask detection in smart cities using deep and transfer learning: lessons learned from the covid-19 pandemic,'' \emph{Systems}, vol.~11, no.~2, p. 107, 2023.

\bibitem{ref3}
S.~S. Mahdi, G.~Battineni, M.~Khawaja, R.~Allana, M.~K. Siddiqui, and D.~Agha, ``How does artificial intelligence impact digital healthcare initiatives? a review of ai applications in dental healthcare,'' \emph{International Journal of Information Management Data Insights}, vol.~3, no.~1, p. 100144, 2023.

\bibitem{ref4}
C.~T. Okolo, ``Optimizing human-centered ai for healthcare in the global south,'' \emph{Patterns}, vol.~3, no.~2, 2022.

\bibitem{ref5}
M.~L. Marinovich, E.~Wylie, W.~Lotter, H.~Lund, A.~Waddell, C.~Madeley, G.~Pereira, and N.~Houssami, ``Artificial intelligence (ai) for breast cancer screening: Breastscreen population-based cohort study of cancer detection,'' \emph{Ebiomedicine}, vol.~90, 2023.

\bibitem{ref6}
R.~L. Siegel, K.~D. Miller, H.~E. Fuchs, A.~Jemal \emph{et~al.}, ``Cancer statistics, 2021,'' \emph{Ca Cancer J Clin}, vol.~71, no.~1, pp. 7--33, 2021.

\bibitem{boughorbel2022predicting}
S.~Boughorbel, Y.~Himeur, H.~E. Salman \emph{et~al.}, \emph{Applications of Machine Learning for Predicting Heart Failure}, 2022, pp. 171--188.

\bibitem{ref8}
N.~Zhang, Y.-X. Cai, Y.-Y. Wang, Y.-T. Tian, X.-L. Wang, and B.~Badami, ``Skin cancer diagnosis based on optimized convolutional neural network,'' \emph{Artificial intelligence in medicine}, vol. 102, p. 101756, 2020.

\bibitem{hamza2023hybrid}
A.~Hamza, B.~Lekouaghet, and Y.~Himeur, ``Hybrid whale-mud-ring optimization for precise color skin cancer image segmentation,'' in \emph{2022 6th International Conference on Signal Processing and Information Security (ICSPIS)}.\hskip 1em plus 0.5em minus 0.4em\relax IEEE, 2023, pp. 1--6.

\bibitem{sayed2022deep}
A.~N. Sayed, Y.~Himeur, and F.~Bensaali, ``Deep and transfer learning for building occupancy detection: A review and comparative analysis,'' \emph{Engineering Applications of Artificial Intelligence}, vol. 115, p. 105254, 2022.

\bibitem{sohail4348272deep}
S.~S. Sohail, Y.~Himeur, A.~Amira, F.~Fadli, W.~Mansoor, S.~Atalla, and A.~Copiaco, ``Deep transfer learning for 3d point cloud understanding: A comprehensive survey,'' \emph{Available at SSRN 4348272}.

\bibitem{ref9}
Y.~Zheng, C.~Li, X.~Zhou, H.~Chen, H.~Xu, Y.~Li, H.~Zhang, X.~Li, H.~Sun, X.~Huang \emph{et~al.}, ``Application of transfer learning and ensemble learning in image-level classification for breast histopathology,'' \emph{Intelligent Medicine}, vol.~3, no.~02, pp. 115--128, 2023.

\bibitem{copiaco2022exploring}
A.~Copiaco, Y.~Himeur, A.~Amira, W.~Mansoor, F.~Fadli, and S.~Atalla, ``Exploring deep time-series imaging for anomaly detection of building energy consumption,'' in \emph{2022 IEEE Asia-Pacific Conference on Computer Science and Data Engineering (CSDE)}.\hskip 1em plus 0.5em minus 0.4em\relax IEEE, 2022, pp. 1--5.

\bibitem{ref10}
R.~Luo and T.~Bocklitz, ``A systematic study of transfer learning for colorectal cancer detection,'' \emph{Informatics in Medicine Unlocked}, p. 101292, 2023.

\bibitem{ref11}
K.~M. Hosny, M.~A. Kassem, and M.~M. Foaud, ``Skin cancer classification using deep learning and transfer learning,'' in \emph{2018 9th Cairo international biomedical engineering conference (CIBEC)}.\hskip 1em plus 0.5em minus 0.4em\relax IEEE, 2018, pp. 90--93.

\bibitem{ref12}
V.~Kumari and R.~Ghosh, ``A magnification-independent method for breast cancer classification using transfer learning,'' \emph{Healthcare Analytics}, p. 100207, 2023.

\bibitem{ref13}
L.~T. Duong, N.~H. Le, T.~B. Tran, V.~M. Ngo, and P.~T. Nguyen, ``Detection of tuberculosis from chest x-ray images: Boosting the performance with vision transformer and transfer learning,'' \emph{Expert Systems with Applications}, vol. 184, p. 115519, 2021.

\bibitem{ref14}
U.~Zidan, M.~M. Gaber, and M.~M. Abdelsamea, ``Swincup: Cascaded swin transformer for histopathological structures segmentation in colorectal cancer,'' \emph{Expert Systems with Applications}, vol. 216, p. 119452, 2023.

\bibitem{ref15}
M.~Springenberg, A.~Frommholz, M.~Wenzel, E.~Weicken, J.~Ma, and N.~Strodthoff, ``From modern cnns to vision transformers: Assessing the performance, robustness, and classification strategies of deep learning models in histopathology,'' \emph{Medical Image Analysis}, vol.~87, p. 102809, 2023.

\bibitem{ref16}
G.~Ayana and S.-W. Choe, ``Buvitnet: Breast ultrasound detection via vision transformers,'' \emph{Diagnostics}, vol.~12, no.~11, p. 2654, 2022.

\bibitem{ref17}
M.~Usman, T.~Zia, and A.~Tariq, ``Analyzing transfer learning of vision transformers for interpreting chest radiography,'' \emph{Journal of digital imaging}, vol.~35, no.~6, pp. 1445--1462, 2022.

\bibitem{ref18}
M.~N. Islam, M.~Hasan, M.~K. Hossain, M.~G.~R. Alam, M.~Z. Uddin, and A.~Soylu, ``Vision transformer and explainable transfer learning models for auto detection of kidney cyst, stone and tumor from ct-radiography,'' \emph{Scientific Reports}, vol.~12, no.~1, p. 11440, 2022.

\bibitem{ref19}
S.~Hossain, A.~Chakrabarty, T.~R. Gadekallu, M.~Alazab, and M.~J. Piran, ``Vision transformers, ensemble model, and transfer learning leveraging explainable ai for brain tumor detection and classification,'' \emph{IEEE Journal of Biomedical and Health Informatics}, 2023.

\bibitem{ref20}
A.~Dosovitskiy, L.~Beyer, A.~Kolesnikov, D.~Weissenborn, X.~Zhai, T.~Unterthiner, M.~Dehghani, M.~Minderer, G.~Heigold, S.~Gelly \emph{et~al.}, ``An image is worth 16x16 words: Transformers for image recognition at scale,'' \emph{arXiv preprint arXiv:2010.11929}, 2020.

\bibitem{ref21}
A.~A. Borkowski, M.~M. Bui, L.~B. Thomas, C.~P. Wilson, L.~A. DeLand, and S.~M. Mastorides, ``Lung and colon cancer histopathological image dataset (lc25000),'' \emph{arXiv preprint arXiv:1912.12142}, 2019.

\bibitem{ref22}
J.~Cheng, W.~Huang, S.~Cao, R.~Yang, W.~Yang, Z.~Yun, Z.~Wang, and Q.~Feng, ``Enhanced performance of brain tumor classification via tumor region augmentation and partition,'' \emph{PloS one}, vol.~10, no.~10, p. e0140381, 2015.

\bibitem{ref23}
J.~Cheng, W.~Yang, M.~Huang, W.~Huang, J.~Jiang, Y.~Zhou, R.~Yang, J.~Zhao, Y.~Feng, Q.~Feng \emph{et~al.}, ``Retrieval of brain tumors by adaptive spatial pooling and fisher vector representation,'' \emph{PloS one}, vol.~11, no.~6, p. e0157112, 2016.

\bibitem{ref24}
P.~Tschandl, C.~Rinner, Z.~Apalla, G.~Argenziano, N.~Codella, A.~Halpern, M.~Janda, A.~Lallas, C.~Longo, J.~Malvehy \emph{et~al.}, ``Human--computer collaboration for skin cancer recognition,'' \emph{Nature Medicine}, vol.~26, no.~8, pp. 1229--1234, 2020.

\bibitem{ref25}
R.~S. Lee, F.~Gimenez, A.~Hoogi, K.~K. Miyake, M.~Gorovoy, and D.~L. Rubin, ``A curated mammography data set for use in computer-aided detection and diagnosis research,'' \emph{Scientific data}, vol.~4, no.~1, pp. 1--9, 2017.

\end{thebibliography}
\end{document}